\documentclass[amssymb,showpacs,amsmath,amssymb,preprint,nofootinbib]{revtex4}
\usepackage{color}
\usepackage{graphicx,bm}

\newcommand{\vect}[1]{\boldsymbol{#1}}

\begin{document}

\title{Gravitational waves from dark matter collapse in a star}

\author{Yasunari Kurita}
\affiliation{Kanagawa Institute of Technology, Atsugi 243-0292, Japan}

\author{Hiroyuki Nakano}
\affiliation{Department of Physics, Kyoto University, Kyoto 606-8502, Japan}

\begin{abstract}
We investigate the collapse of clusters of weakly interacting massive particles (WIMPs) in the core of a Sun-like star and the possible formation of mini-black holes and the emission of gravitational waves.
When the number of WIMPs is small, thermal pressure balances the WIMP cluster's self gravity.
If the number of WIMPs is larger than a critical number, 
thermal pressure cannot balance gravity and the cluster contracts.
If WIMPs are collisionless and bosonic, the cluster collapses directly to form a mini-black hole. 
For fermionic WIMPs, 
the cluster contracts until it is sustained by Fermi pressure, forming a small compact object.
If the fermionic WIMP mass is smaller than $4\times 10^2$ GeV, the radius of the compact object is larger than its Schwarzschild radius and Fermi pressure temporally sustains its self gravity, halting the formation of a black hole. 
If the fermionic WIMP mass is larger than $4\times 10^2$ GeV, the radius is smaller than its Schwarzschild radius and the compact object becomes a mini-black hole.
If the WIMP mass is 1 TeV, the size of the black hole will be approximately 2.5 cm and 
ultra high frequency gravitational waves will be emitted during black hole formation.
The central frequency $f_c$ of ringdown gravitational waves emitted from the black hole will be approximately 2 GHz. 
To detect the ringdown gravitational waves, the detector's noise must be below $\sqrt{S_h(f_c)}\approx 10^{-30}/\sqrt{\rm Hz}$. 
\end{abstract}


\maketitle

\section{Introduction}
\label{sec:intro}

In the near future, it is expected that gravitational waves (GWs) will be directly observed by the second generation of ground-based detectors, 
KAGRA~~\cite{Somiya:2011np,Aso:2013eba}, Advanced LIGO~\cite{TheLIGOScientific:2014jea}, Advanced Virgo~\cite{TheVirgo:2014hva}, GEO-HF~\cite{2014CQGra..31v4002A}, and others.
The GW observations will enrich our understanding of gravity and open a new window to understanding the universe.
The main targets of these detectors are GWs from  
coalescences of neutron stars or black holes (BHs) and supernovae of massive stars. 
To detect them, ground based detectors are sensitive in the frequency range between 10 Hz and 10 kHz.
However, there will be GWs of much higher frequency that are not detectable by the ground-based detectors,
such as background GWs at ultra high frequencies~\cite{Giovannini:1999bh,Giovannini:1998bp,Riazuelo:2000fc,Tashiro:2003qp,Easther:2006gt,GarciaBellido:2007dg,Dufaux:2007pt} and GWs from coalescences of binary primordial BHs~\cite{Nakamura:1997sm,Ioka:1998nz}.
To detect such ultra high frequency GWs, tabletop-sized detectors that are sensitive at frequencies of approximately 100 MHz have been discussed and developed~\cite{Akutsu:2008zza,Akutsu:2008qv,Cruise:2012zz,Goryachev:2014yra,Nishizawa:2007tn}.
If another source of ultra high frequency GWs is discovered, it will further encourage the development of tabletop-sized detectors.

Dark matter (DM) is one of the greatest mysteries in modern physics. DM is believed to exist in the universe but has not been detected. 
An attractive solution is that DM is composed of weakly interacting massive particles (WIMPs) that interact only through gravity and the weak force. 
We can consider that WIMPs would be captured by a star through WIMP--nucleon scattering~\cite{Press:1985ug}, and successive collisions with nucleons in the star would thermalize them to form a cluster of WIMPs.
If the host star is in a region of high DM density and the self annihilation of WIMPs is prohibited, as is the case for asymmetric DM~\cite{Zurek:2013wia,Petraki:2013wwa}, a huge number of WIMPs will be accreted.
Then, a cluster of WIMPs may become a self-gravitating system and finally collapse to a BH~\cite{Goldman:1989nd,Gould:1989gw}, which may lead to the destruction of the host star.
This implies that the existence of old stars in the universe can provide constraints on the WIMP--nucleon cross section, as studied by many researchers~\cite{Bertone:2007ae,deLavallaz:2010wp,Kouvaris:2010vv,McDermott:2011jp,Kouvaris:2010jy,Kouvaris:2011fi,Kouvaris:2011gb,Guver:2012ba,
Bell:2013xk,Bramante:2013hn,Zheng:2014fya}.
The existence of a DM condensate in a star will give rise to various phenomena in the star~\cite{Kouvaris:2007ay,Li:2012qf,Tachibana:2013iva,Kouvaris:2014rja,Bramante:2015dfa,Bramante:2015cua,Bertoni:2013bsa}.

From DM collapse in a star, the emission of GWs with ultra high frequency can be expected, because the size of the BH will be very small. For example, it will be of atomic size in a neutron star~\cite{McDermott:2011jp} and the size of a coin in a Sun-like star~\cite{Kouvaris:2010jy}.
Therefore, GWs from such a collapse might be a target of tabletop-sized detectors.
In this paper, we consider accumulation of asymmetric DM which has effectively no annihilation, and estimate GWs from such mini-BH formations.
For this purpose, it is necessary to evaluate the mass of the mini-BH which is determined by the mass and number of WIMPs,
and hence we investigate the critical number of WIMPs needed for gravitational contraction, using a slightly different method from previous studies~\cite{Kouvaris:2010jy,Bramante:2015dfa,Bramante:2014zca}.
As is well-known, no GWs are emitted from a spherically symmetric collapse. Rotation and deformation are important for GW emission. Stars are usually rotating and therefore a cluster of WIMPs in thermal equilibrium with the host star co-rotates with the star.
Hence, we discuss the rotational effect and find that the rotation may halt direct BH formation from a gravitational collapse due to the centrifugal potential. From this analysis, we obtain a typical rotational parameter for mini-BHs and estimate GWs from the DM collapse.

The paper is organized as follows:
In the next section, we discuss the condition for gravitational collapse using the virial theorem.
Section~\ref{sec:rotation} is devoted to a study of the rotational effect for the collapse.
In section~\ref{sec:accretion}, we estimate the number of WIMPs captured by a star during the lifetime of the star. 
Then in section~\ref{sec:GW}, we discuss GWs emitted from a collapse and obtain the typical frequency and amplitude.
In section~\ref{sec:summary}, we summarize our results.
Appendix~\ref{app:self} gives a brief derivation of the potential term due to self gravity, which is used in the analysis in section~\ref{sec:GC}. 
In appendix~\ref{app:f}, we reevaluate the probability for at least one WIMP--nucleon collision to occur inside a star, which was originally calculated by Press and Spergel~\cite{Press:1985ug}.

In this paper, the speed of light and the Boltzmann constant are set to be one, for simplicity.

\section{condition for gravitational collapse}
\label{sec:GC}

If a low-energy WIMP collides with a nucleon in a star, the WIMP loses its energy and becomes bound by the gravity of the star.
After many WIMPs have accumulated inside the star, they collide with nuclei and form a cluster that is in thermal equilibrium with the core of the star.
In this stage, thermal pressure balances the gravity of the star as well as the self gravity, and the system is in virial equilibrium.
Then, the majority of WIMPs are concentrated within the thermal radius.
As further WIMPs are captured, 
the DM cluster is dominated by the self gravity, and the thermal pressure cannot sustain the gravity
and the cluster starts to contract. 
We estimate the number of WIMPs necessary for gravitational contraction below.

The law of equipartition of energy and the virial theorem give an equation for the radius $r$:
\begin{eqnarray}
\frac{3}{2} T = \frac{GNm^2}{4\sqrt{2} r} + \frac{GmM(r)}{2r} \,,
\label{eq:virial1}
\end{eqnarray}
where $T$ is temperature, $G$ is the gravitational constant, 
$N$ is the number of WIMPs, and $m$ is the WIMP mass. 
$M(r)$ is the mass of the star within a radius $r$,
\begin{eqnarray}
M(r) = \rho_c \frac{4\pi r^3}{3} \,,
\end{eqnarray}
where $\rho_c$ is the core density and is assumed to be a constant.
The first term of the right hand side in Eq.~(\ref{eq:virial1}) arises from the potential of self gravity. We present a brief derivation in Appendix~\ref{app:self}. 
Equation~(\ref{eq:virial1}) is a cubic equation for $r$, and if $N$ is smaller than $N_c$ defined as
\begin{eqnarray}
N_c :=\frac{2\sqrt{6}}{\sqrt{\pi \rho_c} } \left( \frac{T}{G} \right)^{3/2} \frac{1}{m^{5/2}} \,,
\label{eq:Nc}
\end{eqnarray}
it has two positive solutions.
The larger positive solution can be interpreted as the thermal radius since, in the limit $N\to 0$, it tends towards to the thermal radius without the self gravity,
\begin{eqnarray}
r_{\rm th}= \left( \frac{9T}{4\pi G\rho_c m}\right)^{1/2} \,.
\end{eqnarray}
When $N>N_c$, the positive solutions disappear. This means that there is no positive radius satisfying the virial equilibrium between thermal pressure and gravity,
which implies that the thermal pressure cannot sustain the gravity and the cluster starts to contract.  
The number $N_c$ is the critical number necessary for the gravitational contraction.

Using typical values of a Sun-like star, $T=1.5\times 10^7$~K and $\rho_c =150~$g/cm$^3$, 
the critical number is estimated as
\begin{eqnarray}
N_c 
\simeq 9.5\times 10^{48} \left(\frac{\rm 150g/cm^3}{\rho_c} \right)^{1/2}
\left( \frac{T}{\rm 1.5\times 10^7 K} \right)^{3/2}
 \left(\frac{\rm TeV}{m} \right)^{5/2} \,.
\label{eq:Nc-sun}
\end{eqnarray}
For a neutron star, since the core density is much higher than that of the Sun and its temperature is lower, 
the critical number is much smaller and is estimated as
\begin{eqnarray}
N_c \simeq 2\times 10^{39} \left(\frac{\rm 10^{15}g/cm^3}{\rho_c} \right)^{1/2}
\left( \frac{T}{\rm 10^5 K} \right)^{3/2}
 \left(\frac{\rm TeV}{m} \right)^{5/2} \,.
\end{eqnarray}
The mass of a mini-BH after the gravitational contraction can be estimated as $M\simeq m N_c$. Thus a mini-BH in a Sun-like star will be much larger than that in a neutron star, 
and GWs emitted from a collapse in a Sun-like star will be more detectable.
Therefore, we consider a Sun-like star as the host star in the rest of this paper.

At $N=N_c$, equation~(\ref{eq:virial1}) has a double root, which is the final thermal radius just before the gravitational contraction.
For example, the radius is evaluated as $r_{\rm th} = 1.7\times 10^8$~cm for $m=1$~TeV, 
and we consider this as a typical value of the final thermal radius in the next section.

It should be noted that the critical number $N_c$ for a Sun-like star is larger than 
the Chandrasekhar limit for both bosonic particles (see, e.g., Ref.~\cite{McDermott:2011jp})
\begin{eqnarray}
N_{\rm Cha}^B\simeq \left(\frac{m_{pl}}{m} \right)^2 \simeq 1.5\times 10^{32} \left(\frac{\rm TeV}{m}\right)^2 \,,
\end{eqnarray}
and fermionic particles (see, e.g., Ref.~\cite{Kouvaris:2010jy})
\begin{eqnarray}
N_{\rm Cha}^F=\left(\frac{9\pi}{4} \right)^{1/2} \left(\frac{m_{pl}}{m} \right)^3
\simeq 5\times 10^{48} \left(\frac{\rm TeV}{m}\right)^{3} \,.
\end{eqnarray}
We also note that $N_c$ does not depend on whether WIMPs are bosonic or fermionic.
For both bosonic and fermionic WIMPs, the number necessary for gravitational contraction is given by Eq.~(\ref{eq:Nc}).
If WIMPs are bosonic, the DM cluster starts to contract down to the size of a BH as $N$ exceeds $N_c$, forming a BH in the host star.
On the other hand, if WIMP is fermionic, 
the Fermi pressure may stop the collapse and sustain the cluster, forming a compact object.

While we may consider that a cluster collapses to a BH when the number of WIMPs exceeds the Chandrasekhar limit, 
it should be noted that the Chandrasekhar limit gives an upper limit for relativistic particles ($p\gg m$), not for non-relativistic particles.
In our case, whether WIMPs are relativistic or not is not obvious and we should therefore investigate the radius of the compact object.

\begin{figure}
\includegraphics[width=0.49\textwidth]{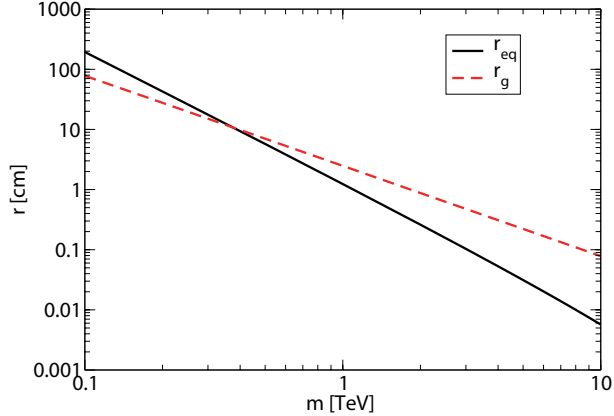}
\caption{Equilibrium radius $r_{\rm eq}$ (black solid line) and 
Schwarzschild radius $r_g$ (red dashed line) of functions of WIMP mass. 
The two radii cross at around $4\times 10^2$~GeV.}
\label{fig:radius}
\end{figure}

In order to estimate the radius of the compact object, 
we consider the virial theorem with a kinetic energy composed of the Fermi momentum,
\begin{eqnarray}
p_{\rm F} = \left(\frac{9\pi}{4} \right)^{1/3} \frac{N^{1/3}}{r} \,.
\end{eqnarray}
Then, the radius $r$ should satisfy the relation:
\begin{eqnarray}
\sqrt{m^2 + p_{\rm F}^2}-m = \frac{GNm^2}{4\sqrt{2} r} + \frac{GmM(r)}{2r} \,.
\label{eq:virial2}
\end{eqnarray}
In the left hand side of this equation 
we use the general expression for the kinetic energy to ensure that 
it is valid for both the non-relativistic (l.h.s. $\sim~p_{\rm F}^2/2m$) and relativistic (l.h.s. $\sim~p_{\rm F}$) cases.
Equation~(\ref{eq:virial2}) can be solved numerically and is found to have one positive solution 
much smaller than $r_{\rm th}$ for $N>N_c$. 
This is an equilibrium radius, $r_{\rm eq}$, at which the Fermi pressure sustains the self gravity.
We might think that the existence of $r_{\rm eq}$ implies that the Fermi pressure
stops the collapse and the cluster becomes a compact object. 
Here, it is valuable to compare $r_{\rm eq}$ with the Schwarzschild radius~\footnote{In this comparison, we ignore the rotational effect and consider the radius of a non-rotating BH, for simplicity.}:
\begin{eqnarray}
r_g = 2GmN_c \,.
\end{eqnarray}
The result is shown in Fig.~\ref{fig:radius}, and is summarized in Table~\ref{tab:r_eq_r_g} for some typical values of $m$. 

\begin{table}
\caption{
Critical number for gravitational contraction $N_c$, equilibrium radius $r_{\rm eq}$ and Schwarzschild radius $r_g$ for typical values of $m$.}
\label{tab:r_eq_r_g}
\begin{tabular}{|c|c|c|c|}
\hline\hline
\ $m$ \ & \ $N_c$ \  & \  $r_{\rm eq}$ [cm] \  & \ $r_g$ [cm] \  \\
\hline 
\ 100 GeV &\  $3.015\times 10^{51}$ &\  $194.1 $&\  $78 $ \\
\hline
\ 375 GeV &\  $1.107\times 10^{50}$ &\  $10.79 $&\  $10.79 $ \\
\hline
\ 1 TeV &\  $9.536\times 10^{48}$ &\  $1.235 $&\  $2.479 $ \\
\hline
\ 10 TeV &\  $3.015\times 10^{46}$ &\  $5.70\times 10^{-3} $&\  $7.84\times 10^{-2} $ \\
\hline\hline
\end{tabular}
\end{table}

As shown in Fig.~\ref{fig:radius} and Table~\ref{tab:r_eq_r_g}, 
$r_{\rm eq}$ is larger than $r_g$ for $m \lesssim 4\times 10^2$~GeV
and smaller for $m \gtrsim 4\times 10^2$~GeV.
Therefore, a compact object supported by Fermi pressure will occur
when $m \lesssim 4\times 10^2$~GeV. 
On the other hand, for $m \gtrsim 4\times 10^2$~GeV, the compact object is contained in 
a region with a Schwarzschild radius $r_g$, being a BH.
Thus, whether or not the cluster collapses to a BH depends on the WIMP mass.

For $m\lesssim 4\times 10^2$~GeV, 
collapse of the cluster does not lead formation of a BH directly but of a DM compact object, 
and therefore significant GW from the collapse of the DM cluster cannot be expected.
After the formation of the compact object, further WIMPs continue to fall on it 
and it eventually becomes so massive that it forms a BH.  
This situation is similar to the gravitational collapse of a rotating neutron star to a BH, 
for which Baiotti et al.~\cite{Baiotti:2005vi} performed numerical simulations and calculated the GW emissions. 
They found that the amplitude of the GWs is not large and the total energy of the emitted GW is 
approximately $10^{-6}$ times mass of the compact object.
Therefore, significant GW emissions also from the collapse of the compact object cannot be expected.

In this paper, we are interested in GW emissions due to the BH formation. 
Therefore, we concentrate our discussion in the following sections on the cases of fermionic WIMPs with $m \gtrsim 4\times 10^2$~GeV
and bosonic WIMPs.
For $m\gtrsim 10$~TeV, the size of the final BH is smaller than $10^{-1}$~cm and 
the frequency of the GWs will be too high to be detectable. 
Also, the GW amplitude from a tiny BH will be too small. 
Hence, we mainly consider the case of $m=1$~TeV.

The detailed contraction mechanism for the cluster strongly depends on
the self interaction between WIMPs or the equation of state for the DM.
However, we do not have any obvious evidence of what it should be,
and hence we assume that WIMPs are collisionless or that self interactions between WIMPs can be negligible as the simplest case.
In this case, after the gravity of the cluster overcomes the thermodynamic pressure, 
there is no force to counteract gravity except for the Fermi pressure,
and the DM cluster contracts on a dynamical timescale.
The contraction can be thought of as a gravitational collapse.

\section{Rotational effect}
\label{sec:rotation}

As a star usually rotates, it is natural to think that a DM cluster in equilibrium
with the star co-rotates with the angular velocity of the star.
As the cluster contracts, it rotates faster due to angular momentum conservation,
which strengthens the centrifugal force.
A strong centrifugal force may halt the gravitational collapse and the formation of a BH,
and therefore we should consider the rotational effect.

The final thermal radius is approximately $r_{\rm th}\simeq 2\times 10^8$~cm for $m=1$~TeV, and is much smaller than the radius of the host (Sun-like) star, which is roughly $0.01R_{\odot}$.
Hence, the DM cluster exists within the core of the star and therefore the rotational period of the core is important
as the initial condition for the collapse.
Though the rotational period of the surface layer or the convection zone of the Sun is known to be from $25$ to $35$~days, that of the core is unknown~\cite{EffDarwich:2012vd,Korzennik}.
We assume that the period of the core is from $20$ to $40$~days. 

\begin{figure}[!ht]
\includegraphics[width=0.49\textwidth]{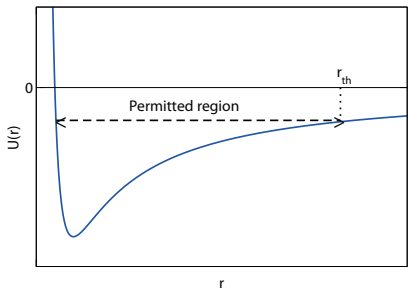}
\caption{Schematic figure for the effective potential $U(r)$. The permitted region for a particle is shown. $r_{\rm th}$ denotes the thermal radius.}
\label{fig:U}
\end{figure}

We briefly 
consider a particle co-rotating in a circular orbit at the thermal radius $r_{\rm th}$ just before the collapse in the manner of classical mechanics. 
The typical effective potential for such a particle is written as
\begin{eqnarray}
U(r)=\frac{L^2}{2mr^2} - G\frac{m^2 N}{r} \,,
\label{eq:Ur}
\end{eqnarray}
where $L$ is the angular momentum of the particle. 
The minimum radius $r_{\rm min}$ that the particle can reach satisfies 
\begin{eqnarray}
U(r_{\rm min}) = U(r_{\rm th}) \,.
\end{eqnarray}
Fig.~\ref{fig:U} shows a schematic figure for $U(r)$ and the permitted region for the particle.
We obtain values of $r_{\rm min}$ for various periods, 
and summarize them in Table~\ref{tab:r_min}.

\begin{table}[!ht]
\caption{The minimum radius $r_{\rm min}$ that a particle can reach,
for various periods of the core.}
\label{tab:r_min}
\begin{tabular}{|c|c|c|c|c|c|}
\hline\hline
\ period of the core ($T_{\rm core}$) \ & \ 20~days \  & \ 25~days \  & \ 30~days \ & \ 35~days \  & \  40~days\  \\
\hline 
\ (angular velocity)/$2\pi$ \ & \ 579~nHz\ & \ 463~nHz \  & \ 386~nHz \  & \ 331~nHz \ & \ 289~nHz  \  \\
\hline
\ $r_{\rm min}$ \ & \ 4.94~cm \  & \ 3.16~cm \  & \ 2.19~cm \ & \ 1.61~cm \  & \  1.23~cm\  \\
\hline\hline
\end{tabular}
\end{table}

For stars having a relatively short rotation period,
$r_{\rm min}$ will be larger than the Schwarzschild radius $r_g \simeq 2.5$~cm.
Therefore, this naive analysis implies that the rotation can halt BH formation and
the cluster `bounces' at $r_{\rm min}$ due to the centrifugal potential. 
After the bounce, the cluster repeats a series of expansions and contractions for a while, entering a radial oscillation phase.
At the bounce, we expect that the cluster has a highly deformed shape with a fast rotation in a strong gravitational field, 
and hence it will emit GWs with nonzero angular momentum,
decreasing the angular momentum of the DM cluster.
For this radial oscillation phase, successive collisions with nuclei
also diminish the angular momentum since a cluster with radius $r<r_{\rm th}$ 
always rotates faster than the core of the star. 
As the angular momentum of the cluster decreases, 
the centrifugal potential reduces in strength.
At some point, the minimum radius becomes less than $r_g$
and the DM cluster collapses to form a mini-BH. 

Therefore, there are two cases: 
1) the cluster starts to contract and collapses to a BH directly, 
or 2) it collapses after the radial oscillation phase.
In both cases, a BH will eventually be formed.

Concerning the dynamical evolution of a highly deformed cluster, 
a numerical study on neutron stars by Giacomazzo et al.~\cite{Giacomazzo:2011cv}
has shown that an artificial reduction in the pressure causes
the evolution of a nonaxisymmetric instability in a highly rotating ``supra-Kerr'' model,
differing from a ``sub-Kerr'' model where a neutron star collapses promptly to a rotating BH.
Therefore, nonaxisymmetric instability may also arise in the collapse of highly rotating DM cluster.
To clarify this, we need to simulate the dynamical evolution of WIMP clusters.
In this paper, we have assumed that a WIMP is collisionless,
and the numerical results obtained by Shibata~\cite{Shibata:1999va} (and references therein) are useful for understanding BH formation from collisionless particles.

\section{Accretion of WIMPs onto a star} 
\label{sec:accretion}

For a self-gravitating DM cluster to form, it is necessary for more than $N_c$ WIMPs to assemble.
This raises the question as to whether it is possible to gather such a huge number of particles. 
In this section, we briefly estimate the number of WIMPs that can be accumulated during the lifetime of the host star.
In order for a huge number of particles to accumulate, regions of extremely high DM density are required.
For this reason, we consider a Sun-like star in the vicinity of the galactic center as the host star.
Though the estimation given in this section is based on 
the analysis of Press and Spergel~\cite{Press:1985ug}, the result will be slightly different.

Following Press and Spergel~\cite{Press:1985ug} 
and Kouvaris~\cite{Kouvaris:2007ay}, 
the accretion rate of WIMPs in a star is written as
\begin{eqnarray}
\frac{dN_{\rm acc}}{dt} = 
8\pi^2  \frac{\rho_{\rm dm}}{m}  \left(\frac{3}{2\pi \bar{v}^2}\right)^{3/2}
GMR\ 
 {\rm min}\left( 
\frac{1}{3}\bar{v}^2, E_0
\right)  f \,,
\label{eq:dNaccdt}
\end{eqnarray}
where $\rho_{\rm dm}$ is the local DM mass density,
$\bar{v}$ is the average WIMP velocity,
$M$ and $R$ are the mass and radius of the host star, respectively, 
$E_0$ is the maximum energy per unit mass of a WIMP that can be captured,
min(A,B) is the minimum value between A and B, 
and $f$ is the probability for at least one WIMP--nucleon scattering event to occur inside the star. 
$E_0$ is expressed by the escape velocity of the star $v_{\rm esc}=\sqrt{2GM/R}$ as
\begin{eqnarray}
E_0 = \frac{\alpha}{2} v_{\rm esc}^2 \,, 
\end{eqnarray}
where $\alpha$ is a constant
determined by the condition that $E_0$ is equal to the mean energy loss in a scattering event, denoted by $\overline{\Delta E}$. 
Assuming isotropic scattering, $\overline{\Delta E}$ is calculated as~\cite{Spergel:1984re}
\begin{eqnarray}
\overline{\Delta E} = 2\frac{m m_p}{(m+m_p)^2} E \,,
\end{eqnarray}
where $m_p$ is the proton mass and $E$ is the WIMP unit mass kinetic energy before the collision.
Then, the condition for $\alpha$ is 
\begin{eqnarray}
 2\frac{m m_p}{(m+m_p)^2} E = \frac{\alpha}{2} v_{\rm esc}^2 \,.
\label{eq:for-alpha}
\end{eqnarray}
If the collision occurs at a radius $r$ inside the star, 
$E$ equals the sum of the kinetic energy at infinity and the decrease in the potential energy,
\begin{eqnarray}
E= \frac{\alpha}{2} v_{\rm esc}^2 + \frac{GM(r) }{r} \,.
\end{eqnarray}
Here, we introduce a parameter $\beta$ that satisfies 
\begin{eqnarray}
\frac{GM(r) }{r} = \beta \frac{v_{\rm esc}^2}{2} \,.
\end{eqnarray}
$\beta=1$ at the surface of the star and increases for collision points deeper inside the star. 
Equation~(\ref{eq:for-alpha}) is written using $\beta$ as
\begin{eqnarray}
 2\frac{m m_p}{(m+m_p)^2} \left( 
\frac{\alpha}{2} v_{\rm esc}^2 +  \frac{\beta }{2}v_{\rm esc}^2
\right)  =  \frac{\alpha}{2} v_{\rm esc}^2 \,.
\label{eq:beta-determin}
\end{eqnarray}
The solution to Eq.~(\ref{eq:beta-determin}) is
\begin{eqnarray}
\alpha  =  2\frac{m m_p}{m^2+m_p^2}  \beta \,,
\end{eqnarray}
and when $m\gg m_p$, $\alpha$ is approximately given by
\begin{eqnarray}
\alpha \simeq 2\frac{m_p}{m} \beta \,.  
\end{eqnarray}
As a typical scattering location, we take half the mass radius of the host star.
From the BP04 solar model~\cite{Bahcall:2004fg}, this radius is $r\simeq 0.25R$.
Hence, from the following relation
\begin{eqnarray}
\frac{GM/2}{0.25R} = \beta \frac{v_{\rm esc}^2}{2} = \beta \frac{GM}{R} \,,
\end{eqnarray}
we obtain $\beta=2$, and $\alpha$ is determined as
\begin{eqnarray}
\alpha \simeq 4 \frac{m_p}{m} \sim3.8\times10^{-3} 
\left( \frac{\rm TeV}{m}\right) \,.
\end{eqnarray}
In order to evaluate the minimum in Eq.~(\ref{eq:dNaccdt}), 
we assume that the average velocity is approximately
$\bar{v} \simeq 10^2~{\rm km/s}$.
The escape velocity of a Sun-like star can be evaluated as
\begin{eqnarray}
v_{\rm esc} = \sqrt{\frac{2GM_{\odot}}{R_{\odot}}} \simeq 6.1\times 10^2~{\rm km/s}\,,
\end{eqnarray}
and therefore 
\begin{eqnarray}
 {\rm min}\left( 
\frac{1}{3}\bar{v}^2, \frac{\alpha}{2} v_{\rm esc}^2 
\right)  = \frac{\alpha}{2} v_{\rm esc}^2 
\simeq
\beta \frac{m_p}{m} v_{\rm esc}^2 
\simeq 
4\frac{m_p}{m}  \frac{GM}{R} \,.
\end{eqnarray}
Now, we can write the accretion rate as
\begin{eqnarray}
\frac{dN_{\rm acc}}{dt} = 
\rho_{\rm dm} 
 \left(\frac{3}{2\pi \bar{v}^2}\right)^{3/2}
32 \pi^2 G^2M^2\ 
 \frac{m_p}{m^2} f \,.
\end{eqnarray}

Next, we calculate the probability $f$, which 
was given by Press and Spergel~\cite{Press:1985ug} as
\begin{eqnarray}
f_{\rm PS} = 0.89  \frac{\sigma}{\sigma_{\rm crit}} \,,
\end{eqnarray}
where $\sigma$ is the WIMP--nucleon scattering cross section and $\sigma_{\rm crit}$ is defined by
\begin{eqnarray}
\sigma_{\rm crit} :=  m_p \frac{R^2}{M} = 4.0\times 10^{-36} \left(\frac{R^2}{R_{\odot}^2} \right)
\left( \frac{M_{\odot}}{M} \right)~{\rm cm^2} \,.
\end{eqnarray}
We evaluate $f$ again using the BP04 solar model and
obtain a different result from Press and Spergel~\cite{Press:1985ug}, as
\begin{eqnarray}
f= 2.6 \frac{\sigma}{\sigma_{\rm crit}} \,.
\end{eqnarray}
The calculation of $f$ is given in Appendix~\ref{app:f}.

The cross section $\sigma$ should satisfy the constraint given by DM search experiments such as XENON100~\cite{Aprile:2013doa}, COUPP~\cite{Behnke:2012ys}, SIMPLE~\cite{Felizardo:2011uw}, PICASSO~\cite{Archambault:2012pm},  LUX~\cite{Akerib:2013tjd}, SuperCDMS~\cite{Agnese:2014aze}, and XMASS~\cite{Abe:2013tc}. 
For a recent review, see Ref.\ \cite{Schumann:2015wfa}. 
The constraint on the spin-independent cross section is very strict,
and it would not be realistic for the critical number of WIMPs given by (\ref{eq:Nc-sun}) to accumulate during the lifetime of the host star
for WIMPs with spin-independent interactions.
On the other hand, the constraint on spin-dependent interactions is not so strict, 
and hence we consider asymmetric DM with spin-dependent interactions.
When spin-dependent interactions are considered, 
the WIMP--He$^4$ interaction is suppressed since it has spin zero. 
Hence, He$^4$ in the host star does not contribute to the scattering probability $f$.
Therefore, we may consider only WIMP--proton scattering.
Assuming that the proton mass ratio in the host star is $75\%$, 
we obtain 
\begin{eqnarray}
f= 2.0 \frac{\sigma}{\sigma_{\rm crit}} \,.
\label{eq:f-final}
\end{eqnarray}
The constraint on spin-dependent WIMP-proton scattering is currently understood to be
as~\cite{Schumann:2015wfa}
\begin{eqnarray}
\sigma  < 3\times 10^{-38} \left(\frac{m}{\rm TeV} \right)~{\rm cm^2 } \,.
\end{eqnarray}

At the galactic center, the DM density can be considered to be extremely high, as in the Navarro--Frenk--White DM model~\cite{Navarro:1995iw}, 
\begin{eqnarray}
\rho_{\rm NFL}(r) = \rho_0 \left(\frac{r}{r_s}\right)^{-\alpha} \left( 1+ \frac{r}{r_s}\right)^{-3+\alpha},
\end{eqnarray}
where $\alpha$ is the inner slope for the DM density profile near the galactic center, 
$r_s$ is the scaling radius and $\rho_0$ is the scale density. 
With $\alpha=1.3$, $\rho_0=$0.4 GeV/cm$^3$ and $r_s=20$ kpc, the DM density is more than $10^7$ GeV
within 0.06 pc of the galactic center, 
 and we assume $\rho_{\rm dm} = 10^7~{\rm GeV/cm^3}$.
One might think that constraints from gamma ray emission produced via WIMPs annihilation exclude
such extremely high DM density.
However,  in this paper, we consider DM candidates of asymmetric nature, which have effectively no annihilation, 
and therefore gamma ray production via WIMPs annihilation does not apply in our case. 
The velocity distribution of DM in the galaxy has not been established yet and, 
especially, the average velocity near the galactic center is not known. 
Mao et al.~\cite{Mao:2012hf} has suggested that it might be slower than $220$~km/s in the inner region of the galaxy.  
Hence, we simply assume $ \bar{v} \simeq 1\times 10^2~{\rm km/s}$.

Finally, the number of WIMPs accreted in a Sun-like star near the galactic center
during $10$ billion years is estimated as
\begin{eqnarray}
N_{acc} 
= 2.5 \times 10^{49} 
\left(\frac{\rho_{\rm dm}}{\rm 10^7GeV/cm^3} \right)
\left(\frac{\rm 100km/s}{\bar{v}} \right)^{3}
\left(\frac{M}{M_{\odot}} \right)^3
\left(\frac{R_{\odot}}{R} \right)^{2}
\left(\frac{\rm TeV}{m} \right)^{2}
\left(\frac{\sigma}{\rm 10^{-38} cm^2} \right) \,,
\nonumber \\
\end{eqnarray}
which is sufficiently beyond the critical number given in Eq.~(\ref{eq:Nc}). 
Therefore, under some convenient assumptions, 
it is possible to accumulate more than $N_c$ particles 
in the host star, provoking the gravitational collapse.

After the accretion of WIMPs, successive collisions with protons make WIMPs thermalize.
In order that the DM cluster can actually collapse, 
the time scale of thermalization should be shorter than the lifetime of the host star. 
The time scale depends on the WIMP-proton cross section, and it should be checked.
As discussed in Ref.~\cite{Kouvaris:2010jy}, the thermalization of captured WIMPs can be characterized by two stages.
At the first stage, the trapped WIMPs oscillate in the star's gravitational potential 
and their orbits go beyond the size of the star.
When the WIMP crosses the star, it may collide with protons and lose some energy, 
making the size of WIMP's orbit smaller. 
At the second stage, the orbit of the WIMP is inside of the star,
and successive collisions make it shrink to the thermal radius.
The timescale of these two stage was evaluated in Ref.~\cite{Kouvaris:2010jy}.
The duration of the first stage is
\begin{eqnarray}
t_1 =3\times 10^3 \,{\rm yr} \left(\frac{m}{\rm TeV} \right)^{3/2} \left(\frac{\sigma}{\rm 10^{-38}cm^2} \right)^{-1} \,,
\end{eqnarray}
and the timescale of the second stage is 
\begin{eqnarray}
t_2 = 1.5 \times 10^2 \,{\rm yr} \left(\frac{m}{\rm TeV} \right) \left(\frac{\sigma}{\rm 10^{-38}cm^2} \right)^{-1} \,.
\end{eqnarray}
These time scales are very short compared with those of stellar evolution and WIMP accretion.

\section{gravitational waves}
\label{sec:GW}

In this section, we briefly discuss GWs emitted from the collapse of a DM cluster. 
We extrapolate and apply the results for gravitational collapses of ordinary stars
to clusters of WIMPs in the radial oscillation phase,
based on Ref.~\cite{Fryer:2011zz}.
We also estimate the GWs of the quasinormal modes
associated with BH formation. 
For numerical estimations in this section, we always consider the case $m=1$~TeV.

First, if there is any asymmetry in the radial oscillation phase,
bursts of GWs are emitted at the bounces.
Using Eq.~(3) of Ref.~\cite{Fryer:2011zz}, which is a simple quadrupole formula,
and considering only the velocity term, we estimate the GW amplitude as
\begin{eqnarray}
h \approx \sqrt{\frac{8}{3}} \frac{G m_m}{d} v_m^2  \,,
\end{eqnarray}
where $d$ is the distance to the GW source location
and $v_m$ and $m_m$ denote the velocity and mass of the asymmetric component 
of the collapsing matter, respectively.
Adopting the case with $r_{\rm min}=3.16$~cm in Table~\ref{tab:r_min},
the maximum velocity becomes $v \approx 0.44$ at $r=6.33$~cm.
Assuming that $1\%$ of the mass of the cluster of WIMPs,
i.e., $m_m=0.01m N_c$, contributes to the GW emission,
together with $v_m \approx 0.44$, 
the GW amplitude is obtained as
\begin{eqnarray}
h \approx 1.6 \times 10^{-25} \left(\frac{d}{8{\rm kpc}}\right)^{-1} \,.
\end{eqnarray}

The characteristic GW frequency is calculated as follows.
We assume that the GW burst is emitted around the maximum velocity.
Approximating the effective potential in Eq.~(\ref{eq:Ur}) as
\begin{eqnarray}
U(r) = -\frac{G^2 m^5 N^2}{2 L^2}
+ \frac{G^4 m^{11} N^4}{2 L^6} (r-r_0)^2 \,,
\end{eqnarray}
where $r_0=L^2/(G m^3 N)$ is the location of the potential minimum,
i.e., the maximum velocity, we derive the period of the oscillation.
The inverse of the period gives the characteristic frequency,
\begin{eqnarray}
f  \approx  3.3 \times 10^8 \, {\rm Hz} \,.
\end{eqnarray}
Although this looks reasonable compared with the frequency of the quasinormal mode
discussed later, a detailed study will be required to obtain
a more precise GW frequency and amplitude.
Also, we should note that the assumption on the mass contributed to the GW emission
is a conservative one and leads to the radiated GW energy $E \approx 2 \times 10^{-6}$\,\%
by using Eq.~(\ref{eq:amp_E}) below for one bounce with the characteristic frequency,
but we need numerical simulations to remove this ambiguous assumption.

Next, we consider GWs emitted due to BH formation.
In the nonlinear and dynamical era of BH formation, strong bursts of GWs will be emitted, and
numerical relativistic simulations are required to estimate them. 
However, this is outside of the scope of this paper. 
We focus only on ringdown GWs due to the excitation of quasinormal modes in the final phase of BH formation.
When the period of the core satisfies $T_{\rm core} \lesssim 28$~days,
the collapse undergoes a radial oscillation phase. After a sufficient decrease in angular momentum, 
$r_{\rm min}$ will be smaller than the BH's horizon radius
(for simplicity, we consider the Schwarzschild radius,
i.e., $r_g=2 M_{\rm BH}\approx 2.5$~cm, instead of the radius of the Kerr BH).
When the angular momentum decreases to that corresponding to the case of $T_{\rm core} \approx 28$~days, 
the cluster will collapse to form a BH. 
However, in the case of $T_{\rm core} \gtrsim 28$~days,
a BH will directly form from the DM cluster
without a radial oscillation phase.
For the above situations, we calculate
the amplitude and frequency of the ringdown GWs.

As a rough estimation, 
we treat the cluster of WIMPs as a rigidly rotating ball in classical mechanics.
Considering the angular momentum conservation during the shrinking phase,
we obtain the angular velocity of the BH, 
\begin{eqnarray}
\omega_{\rm BH} = \left( \frac{r_{\rm th}}{r_g} \right)^2 \omega_f \,,
\end{eqnarray}
where $\omega_f$ is defined as follows: when the period of the core is $T_{\rm core} \gtrsim 28$~days, $\omega_f$ is the angular velocity of the core 
and when  $T_{\rm core} \lesssim 28$~days,  $\omega_f = 2\pi /(28~{\rm days})$, 
which is the angular velocity of the core with $T_{\rm core} \approx 28$~days.
Here, we introduce a nondimensional spin parameter
$\chi_{\rm BH} = G M_{\rm BH}\ \omega_{\rm BH}$.
Then, the BH spin is related to the angular velocity of the star's core as 
\begin{eqnarray}
\chi_{\rm BH} \approx \left( \frac{r_{\rm th}}{r_g} \right)^2
G M_{\rm BH} \ \omega_f \,.
\end{eqnarray}
We consider the above $\chi_{\rm BH}$ as the spin parameter
of the BH formed by the gravitational collapse.
Here, $(r_{\rm th}/r_g)^2  \approx 4.7 \times 10^{15}$ is a huge number.
Thus, we obtain the BH spin parameter,
\begin{eqnarray}
\chi_{\rm BH} \approx 
0.5 \left( \frac{T_{\rm core}}{28~{\rm days}} \right)^{-1} \,.
\end{eqnarray}
In the case of $T_{\rm core} \lesssim 28$~days, we assign $T_{\rm core} = 28$~days in the above evaluation.

Based on Ref.~\cite{Berti:2005ys}, we estimate the amplitude
of the ringdown GWs as
\begin{eqnarray}
A \approx \sqrt{\frac{8 GE}{f_c Q}} \,,
\label{eq:amp_E}
\end{eqnarray}
where the central frequency $f_c$ and 
quality factor $Q$ are given as fitting functions,
\begin{eqnarray}
f_c &=& \frac{1}{2\pi G M_{\rm BH}}
\left[ 1.5251 - 1.1568 (1-\chi_{\rm BH})^{0.1292} \right] 
\cr 
&\approx& 3.9 \times 10^{9}
\left[ 1.5251 - 1.1568 (1-\chi_{\rm BH})^{0.1292} \right]
\, {\rm Hz} \,,
\cr
Q &=& 0.7000 + 1.4187 (1-\chi_{\rm BH})^{-0.4990} \,.
\end{eqnarray}
These parameters are related to the real ($f_R$)
and imaginary ($f_I$) parts of the quasinormal modes satisfying 
$f_R = f_c$ and $f_I = - f_c/(2Q)$.

Concerning the estimation of the radiated energy $E$,  numerical simulations are helpful.
Detailed investigation of GWs from the collapse of neutron stars was given by Baiotti et al.~\cite{Baiotti:2007np}, showing that $E$ is approximately $10^{-4}~\%$ of the BH mass for $\chi_{\rm BH} =0.5$. 
However, the collapse of the DM cluster would not be similar to that of neutron stars.
The reason is as follows: Before the collapse, the DM cluster is in thermal equilibrium with the baryonic fluid by successive collisions between protons and WIMPs.  As the thermalization time which is estimated as $10^3$ years (see e.g.,~\cite{Kouvaris:2010jy}) is much longer than the dynamical timescale, the WIMP-proton interaction becomes negligible after the cluster starts to contract. 
Therefore, the thermal pressure which supports the cluster will also be ineffective, which is much different from the case of the neutron stars.
Our setup would be similar to the collapses of rotating stars via sudden pressure depletion considered by Stark and Piran~\cite{Stark:1985da}.
According to the numerical simulation given by Stark and Piran~\cite{Stark:1985da}, 
which preserves an axisymmetric configuration,
we assume that the radiated energy, $E$,
is $0.01\%$ of the BH mass for $\chi_{\rm BH} =0.5$. 
Now, the GW amplitude is calculated as
\begin{eqnarray}
h = \frac{A}{d} \approx 3.2 \times 10^{-24} \left(\frac{d}{8{\rm kpc}}\right)^{-1} \,,
\label{eq:RGh}
\end{eqnarray}
where we have set $\chi_{\rm BH} =0.5$
as the characteristic spin value.
The parameters of the ringdown GWs are $f_c \approx 1.8 \times 10^9$~Hz
and $Q \approx 2.7$. 
In this evaluation, only a small fraction of the BH mass contributes to the radiated GW energy
due to the axisymmetry. 
We may expect larger GW amplitudes if there is any break of the axisymmetry.

This is not the end of the story however.
In practical GW observations, we need to consider the signal-to-noise ratio (SNR).
Using Ref.~\cite{Berti:2005ys} (see also, Ref.~\cite{Flanagan:1997sx}),
the SNR for the above ringdown GWs is evaluated by
\begin{eqnarray}
({\rm SNR}) \approx \frac{\sqrt{5}}{10\pi} \frac{A}{d} \sqrt{\frac{Q}{f_c S_h(f_c)}}
\approx 
8.9 \times 10^{-8} 
\left(\frac{d}{8{\rm kpc}}\right)^{-1}
\left(\frac{\sqrt{S_h(f_c)}}{10^{-22}/\sqrt{\rm Hz}}\right)^{-1}
\,,
\end{eqnarray}
where $S_h(f_c)$ is the noise spectral density of the GW detector
at the central frequency, for which we have used the value given
in Ref.~\cite{Goryachev:2014yra}.
This means that to detect ringdown GWs, 
the detector's noise must be below $\sqrt{S_h(f_c)} \approx 10^{-30}/\sqrt{\rm Hz}$.

Note that in Ref.~\cite{Fryer:2001zw}
various possibilities are suggested for the emission of GWs from the system.
Especially, GWs from bar-mode and fragmentation instabilities
have been estimated for massive stars.
In our situation, these GWs will be radiated in the case of 
$m \lesssim 4\times 10^2$~GeV
where the gravitational collapse does not lead directly to BH formation
and the cluster of WIMPs is supported by Fermi pressure.
A study in this direction is left for future work 
since it is necessary to discuss the growth
of these instabilities numerically.
Also, there is ambiguity in the radiated energy of ringdown GWs.
Since the ringdown efficiency is sensitive to the collapsing mechanism~\cite{Berti:2009kk},
this will require in-depth numerical simulations.

\section{summary and discussion}
\label{sec:summary}

We have calculated the critical number of WIMPs necessary for gravitational collapse.
We found that the mass of a BH increases
when the host star has a lower baryonic density and higher temperature.
This means that a BH formed in a Sun-like star is larger than one in a neutron star, 
and GWs emitted from the former BH are more detectable.
Hence, we have considered a Sun-like star as the host star. 
One of the main results is that if WIMPs are fermions and their mass is smaller than $4\times 10^2$~GeV, 
a gravitational collapse leads to a compact object supported by Fermi pressure, 
and otherwise a BH is formed.

Taking into account the rotational effect, 
we have found that
the centrifugal force may halt BH formation for a while.
For $m=1$~TeV, if the core of the star rotates slowly, 
the DM cluster collapses to a BH directly.
On the other hand, if it rotates with a period of $T_{\rm core}\lesssim  28$~days, 
the centrifugal potential prevents BH formation and the cluster gives way to
a radial oscillation phase.
At each bounce of the oscillation, a fraction of the angular momentum of the cluster is extracted
by GW emission, and it finally collapses to a BH.

For a sufficient number of WIMPs to accumulate for self gravitating in a Sun-like star, 
WIMPs exhibiting spin-dependent interactions should be considered.
Furthermore, since an extremely high DM density is required,
the host star should be in the vicinity of the galactic center.
Therefore, we have set the distance from the GW source to be 8~kpc,
which is used in our estimation of GW amplitudes.

For WIMPs with $m=1$~TeV, two types of GWs from the collapse have been estimated.
At each bounce, GWs with a characteristic frequency of $3\times 10^8$~Hz will be emitted.
After BH formation, quasinormal ringing GWs
will be produced with the central frequency estimated to be $2\times 10^9$~Hz,
while the amplitude has been estimated to be $h \approx 3 \times 10^{-24}$.
To detect ringing GWs, the detector's noise level should be lower than
$\sqrt{S_h(f_c)} \approx 10^{-30}/\sqrt{\rm Hz}$.
This makes the detection of GWs from DM collapses
quite challenging.

We have assumed that WIMPs are collisionless and the gravitational contraction occurs in a dynamical way.
However, self interactions between WIMPs will affect the gravitational contraction and therefore also the GW emission.
Conversely, if GWs from the DM collapse are detected, the details of the self interaction
can be investigated by ultra-high-frequency GW observations. 
Therefore, GW observation might be able to be a tool to investigate these interactions.

In the analysis of GWs, 
we have evaluated the radiated energy of ringdown GWs by adopting the result for an axisymmetric collapse~\cite{Stark:1985da} in which only 0.01\% of the BH mass is converted to GW energy.
However, in actual cases, complete preservation of the axisymmetric configuration cannot be expected, as in the case of neutron stars~\cite{Giacomazzo:2011cv}, and 
it was shown that a much larger fraction of energy is emitted as GWs in nonaxisymmetric collapses (see, e.g.,~\cite{Mroue:2013xna,Healy:2014yta}
for binary BH mergers). 
Therefore, the amplitude of ringdown GWs is expected to be larger than that given in Eq.~(\ref{eq:RGh}).

Recently, Brito, Carodoso, and Okawa~\cite{Brito:2015yga} showed by solving field equations numerically that a DM cluster may stop growing due to gravitational cooling, if the DM is composed of light massive bosonic fields.
Their DM configurations seem to be coherent, like Bose--Einstein condensates, which is not applicable to our case. 
However, their simulation clearly shows the importance of numerical investigation for constructing solutions. 
As the authors stated, other more detailed simulations will be needed.

\section*{Acknowledgements}

We would like to thank Takahiro Tanaka and Chul-Moon Yoo for helpful discussions. 
Y.K. is grateful to Motoi Tachibana for useful discussions.
H.N. acknowledges support by 
the Ministry of Education, Culture, Sports, Science and Technology (MEXT)
Grant-in-Aid for Scientific Research on Innovative Areas,
``New Developments in Astrophysics Through Multi-Messenger Observations
of Gravitational Wave Sources'', No.~24103006.


\appendix

\section{Virial theorem for self gravity}
\label{app:self}

Let us consider a system including $N$ particles 
and the following time derivative:
\begin{eqnarray}
\frac{d}{dt} \sum_{a=1}^N m_a \vect{v}_a \cdot \vect{x}_a
&=& \sum_{a=1}^N m_a \dot{\vect{v}}_a \cdot \vect{x}_a
+ \sum_{a=1}^N m_a \vect{v}_a \cdot \dot{\vect{x}}_a
\cr 
&=& \sum_{a=1}^N \left( \sum_{b=1,b\neq a}^N \frac{G m_a m_b (\vect{x}_b-\vect{x}_a)}{r_{ab}^3} \right) \cdot \vect{x}_a
+ \sum_{a=1}^N m_a v_a^2
\cr
&=& -\frac{1}{2} \sum_{a=1}^N \sum_{b=1,b\neq a}^N \frac{G m_a m_b}{r_{ab}} 
+ \sum_{a=1}^N m_a v_a^2 \,, 
\end{eqnarray}
where $r_{ab}=|\vect{x}_b-\vect{x}_a|^{1/2}$
and $\vect{v}_a=d\vect{x}_a/dt=\dot{\vect{x}}_a$. 
Assuming that $N$ particles have the same mass $m$ and taking the average of this derivative
over a long time, the left hand side of the above equation vanishes and we obtain 
\begin{eqnarray}
0 = -\left< \frac{1}{2} \sum_{a=1}^N \sum_{b=1,b\neq a}^N \frac{G m}{r_{ab}} \right>
+ \left< \sum_{a=1}^N v_a^2 \right> \,.
\end{eqnarray}
Averaging over all particles and considering one typical particle, this equation can be written as
\begin{eqnarray}
0 = - \frac{1}{2} Gm  \frac{1}{N} \sum_{a=1}^N \sum_{b=1,b\neq a}^N   \left<\frac{1}{r_{ab}} \right>_{\rm part. \,\,ave.}
+ \frac{1}{N} \sum_{a=1}^N  \left< v_a^2 \right>_{\rm part. \,\,ave.}
\label{eq:virial-1}
\end{eqnarray}
Now, we consider an approximation,
\begin{eqnarray}
\left< r_{ab} \right>_{\rm part. \,\,ave.}
= \left< \sqrt{ r_b^2 - 2 \vect{x}_b \cdot \vect{x}_a + r_a^2 }\right>_{\rm part. \,\,ave.} 
\approx \sqrt{2} \left< r_a \right>_{\rm part. \,\,ave.}
=: \sqrt{2} r \,, 
\end{eqnarray}
where we have defined the particle averaged radius $r$ in the previous equation.
We introduce the root mean square velocity $\bar{v}$, defined as
\begin{eqnarray}
 \frac{1}{N} \sum_{a=1}^N  \left< v_a^2 \right> =   \overline{v}^2 \,,  
\end{eqnarray}
Eq.~(\ref{eq:virial-1}) leads to
\begin{eqnarray}
\frac{1}{2} m\bar{v}^2 = 
 \frac{G N m^2}{4\sqrt{2}r} \,.
\end{eqnarray}
Adding a contribution from the gravitational potential of the host star, we have
\begin{eqnarray}
\frac{1}{2} m\bar{v}^2 = 
 \frac{G N m^2}{4\sqrt{2}r} 
+ \frac{Gm M (r) }{2r} \,,
\end{eqnarray}
which gives Eq.~(\ref{eq:virial1}).

\section{evaluation of $f$}
\label{app:f}

In this appendix, we calculate the probability $f$ in Eq.~(\ref{eq:dNaccdt}), 
based on the derivation given by Press and Spergel~\cite{Press:1985ug}.
Since the WIMP--nucleon cross section is very small, $f$ can be approximated as 
\begin{eqnarray}
f \approx \left<
\int \frac{\rho_{\rm p} \sigma}{m_{\rm p}} d\ell 
\right>
=  \frac{\sigma}{\sigma_{\rm crit}} \frac{R^3}{M} \left< \int \rho_p \frac{d\ell}{R}  \right> \,,
\end{eqnarray}
where $m_{\rm p}$ and $\rho_{\rm p}$ are 
the proton mass and mass density, respectively.
The brackets mean an average over capturable orbits.
$d\ell$ denotes an infinitesimal arc length along an orbit,
which is written as
\begin{eqnarray}
d\ell 
=\sqrt{\left( \frac{dr}{d\theta}\right)^2 + r^2 } \, d\theta \,,
\end{eqnarray}
where $\theta$ is the angular variable of the orbit
and we set $\theta=0$ at the surface $r=R$.

We introduce the dimensionless variables
\begin{eqnarray}
\hat{J}^2 = \frac{J^2}{GMR} \,, \quad 
\hat{u} = \frac{R}{r} \,, \quad 
\hat{M}(r) = \frac{M(r)}{M} \,, 
\end{eqnarray}
where $J$ denotes the specific angular momentum of the WIMP.
Considering only a low-energy particle ($E\approx 0$), which will be captured by the star,
the law of energy conservation per unit mass is approximately given as
\begin{eqnarray}
\left( \frac{d\hat{u}}{d\theta} \right)^2
+\hat{u}^2 - \frac{2}{\hat{J}^2} \int_0^{\hat{u}} \hat{M}(r) d\hat{u} \approx 0 \,.
\end{eqnarray}
Then, we obtain
\begin{eqnarray}
\frac{d\ell}{R} 
= \frac{1}{\hat{u}^2} \sqrt{\left( \frac{d\hat{u}}{d\theta}\right)^2 + \hat{u}^2 } \, d\theta
= \frac{1}{\hat{u}^2} \sqrt{ \frac{2}{\hat{J}^2} \int_0^{\hat{u}} \hat{M}(r) d\hat{u} } \, d\theta
\,.
\end{eqnarray}
Since $\hat{M}(r) =1$ outside the star ($0\le \hat{u} \le 1$),
we have $\int_0^1 \hat{M}(r) d\hat{u} =1$.
The orbit $\hat{u}(\theta)$ inside the star is determined by the equation of motion,
\begin{eqnarray}
\frac{d^2\hat{u}}{d\theta^2} +\hat{u} -\frac{\hat{M}(r)}{\hat{J}^2} = 0 \,,
\label{eq:EOM-u}
\end{eqnarray}
with the initial conditions 
\begin{eqnarray}
\hat{u}(0)=1 \,, 
\quad \frac{d\hat{u}}{d\theta}\big|_{\theta=0} = \sqrt{\frac{2}{\hat{J}^2}-1} \,.
\end{eqnarray}

\begin{figure}[!ht]
\includegraphics[width=0.49\textwidth]{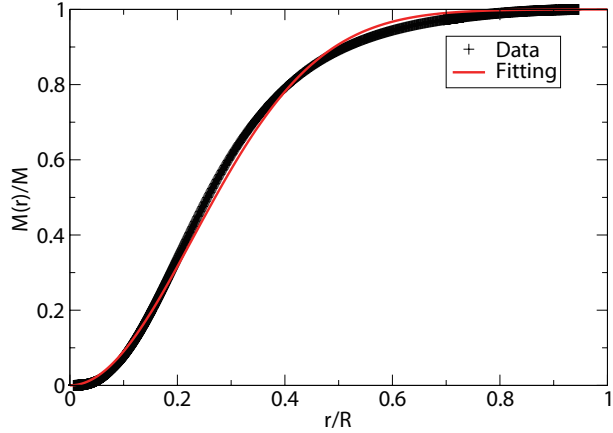}
\caption{Mass fraction of the BP04 solar model~\cite{Bahcall:2004fg}, showing the data (``$+$'' symbols, which run together to form the thick black line) and
the fit using Eq.~(\ref{eq:M-fitting}) (thin red thin line).}
\label{fig:SSM2}
\end{figure}
\begin{figure}[!ht]
\includegraphics[width=0.49\textwidth]{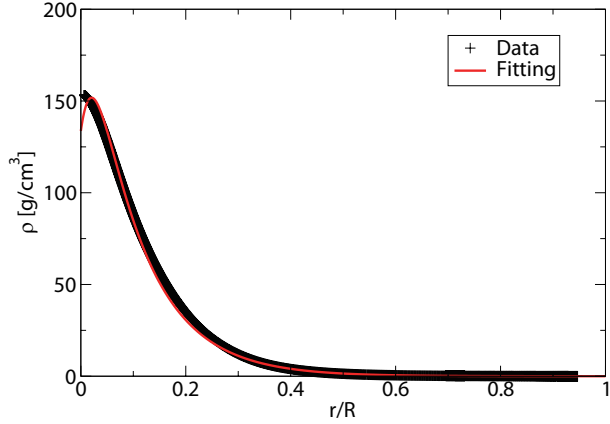}
\caption{Solar density, showing the data (``$+$'' symbols) 
and the fit
using Eq.~(\ref{eq:density-fit}) (thin red line).}
\label{fig:solar-density}
\end{figure}

To solve Eq.~(\ref{eq:EOM-u}), we need an expression for 
the mass fraction $\hat{M}(r)$ in units of solar mass.
In Fig.~\ref{fig:SSM2}, we show the mass fraction given by the BP04 solar model Ref.~\cite{Bahcall:2004fg}
and a function,
\begin{eqnarray}
\hat{M(r)} = 1-\exp\left[-9.5 \left(\frac{r}{R}\right)^2\right] 
= 1-\exp\left[-\frac{9.5}{\hat{u}^2} \right].
\label{eq:M-fitting}
\end{eqnarray}
As shown in Fig.~\ref{fig:SSM2}, the function (\ref{eq:M-fitting}) approximately fits the mass fraction, 
and we adopt it as the fitting function below.
We calculate the integration of $\hat{M}(r)$ with respect to $\hat{u}$ as
\begin{eqnarray}
\int_0^{\hat{u}} \hat{M}(r) d\hat{u}  
&=& \int_0^1  d\hat{u} +\int_1^{\hat{u}} \hat{M}(r) d\hat{u}  
 = 1 + \int_1^{\hat{u}} \hat{M}(r) d\hat{u} 
\cr
&=& 
\hat{u}-\hat{u}\exp\left(-\frac{9.5}{\hat{u}^2} \right) 
-\sqrt{9.5\pi} {\rm Erf}\left( \frac{\sqrt{9.5}}{\hat{u}} \right)
+\left(e^{-9.5}+ \sqrt{9.5\pi} {\rm Erf}(\sqrt{9.5})  \right) \,.
\nonumber \\
\end{eqnarray}
We also need the density of the solar model,
which can be fitted by the following function,
\begin{eqnarray}
\rho_p(\hat{r}) =\frac{ 200(\hat{r}+1.75)^3}{\exp[11.5\hat{r}+1.6]-1} \tanh(18(\hat{r}+0.03))
\,,
\label{eq:density-fit}
\end{eqnarray}
as shown in Fig.~\ref{fig:solar-density}. 
\begin{figure}[!ht]
\includegraphics[width=0.49\textwidth]{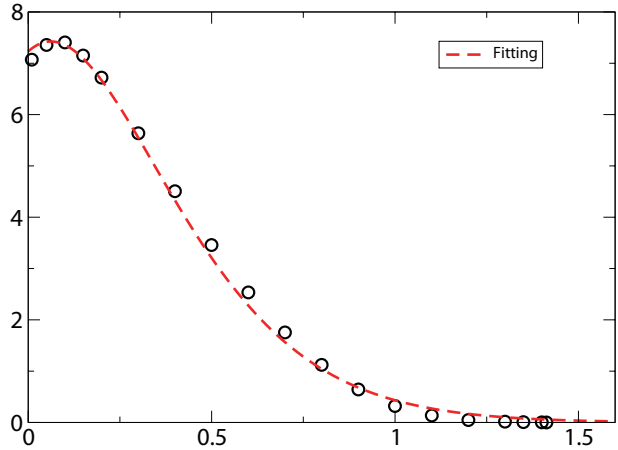}
\caption{Numerical values of integral $(R^3/M) \int \rho_p \,d\ell/R$ for several values of $\hat{J}$ (circles)
and the fit using Eq.~(\ref{eq:f-fitting}) (red dashed line).
The horizontal axis is $\hat{J}$. }
\label{fig:f-integral}
\end{figure}
Using these fitting functions and solving the equation of motion in Eq.~(\ref{eq:EOM-u}),
we perform the integration
\begin{eqnarray}
\frac{R^3}{M} \int \rho_p \frac{d\ell}{R} 
= \frac{R^3}{M} \int d\theta \rho_p   \frac{1}{\hat{u}^{2}} \sqrt{\frac{2}{\hat{J}^2} \int_0^{\hat{u}} \hat{M}(r) d\hat{u} } \,,
\label{int:f-including-rho}
\end{eqnarray}
for several values of $\hat{J}$ numerically. 
The results are shown in Fig.~\ref{fig:f-integral} and can be fitted by a function,
\begin{eqnarray}
\tilde{f}(\hat{J}) = 
0.73(10\hat{J}+4.8)^4 \left(\exp(7.3\hat{J}+4)-1 \right)^{-1} \,.
\label{eq:f-fitting}
\end{eqnarray}
Averaging over the orbits with $E\approx 0$ and $0<J^2<2GM$,  
the integration of $\tilde{f}$ with respect to $\hat{J}$ gives 
\begin{eqnarray}
\frac{1}{\sqrt{2}} \int_0^{\sqrt{2}} \tilde{f}( \hat{J})  d\hat{J}
= 2.6418 \,.
\end{eqnarray}
Finally, we obtain 
\begin{eqnarray}
f= 2.6 \frac{\sigma}{\sigma_{\rm crit}} \,.
\end{eqnarray}



\end{document}